\documentclass[11pt]{article}
\usepackage{amsmath}
\usepackage{graphicx}
\usepackage{amsfonts}
\usepackage{amssymb}
\usepackage{amscd}
\newcommand{\bc}{\begin{center}}
\newcommand{\ec}{\end{center}}
\newcommand{\be}{\begin{equation}}
\newcommand{\ee}{\end{equation}}
\newcommand{\bea}{\begin{eqnarray}}
\newcommand{\eea}{\end{eqnarray}}
\newcommand{\ba}{\begin{array}}
\newcommand{\ea}{\end{array}}
\newcommand{\edc}{\end{document}}

\def\f{\varphi}

\def\g{\gamma}

\def\Om{\Omega}

\def\b{\beta}

\def\s{\sigma}
\def\m{\mu}

\def\l{\lambda}
\def\L{\Lambda}

\def\d{\partial}

\begin{document}
\thispagestyle{empty}
\begin{center}

{\Large {\bf Pirogov-Sinai Theory With  New Contours For Symmetric Models}}\\
\vspace{0.4cm}
{\bf N.N. Ganikhodjaev$^{1,2}$, U.A. Rozikov$^{1,3}$}\\
 $^1$Institute of Math. and Infor. Technol., 29, F.Hodjaev str., 100125, Tashkent,
 Uzbekistan.\\
$^2$International Islamic University Malaysia, P.O. Box 141, 25710, Kuantan, Malaysia.\\
$^3$ School of Math. Sci. GC University, Lahore, Pakistan.\\
 E-mail: nasirgani@yandex.ru, rozikovu@yandex.ru\\[2mm]
\end{center}
\vspace{0.5cm}

{\bf Abstract:} The contour argument was introduced by Peierls for
two dimensional Ising model. Peierls benefited from the particular
symmetries of the Ising model. For non-symmetric models the argument
was developed by Pirogov and Sinai. It is very general and rather
difficult. Intuitively clear that the Peierls argument does work for
any symmetric model. But contours defined in Pirogov-Sinai theory do
not work if one wants to use Peierls argument for more general
symmetric models. We give a new definition of contour which allows
relatively easier prove the main result of the Pirogov-Sinai theory
for symmetric models. Namely, our contours allow us to apply the
classical Peierls argument (with contour removal operation).

(primary);
60K35, 05C05 (secondary).\\[1mm]

\section{Introduction}

    In many systems of interest, low temperature Gibbs measures
are concentrated on configurations which are basically a single
configuration plus a small fraction of small "fluctuations",
also called "defects". The boundaries of these "fluctuations",
define the contours.

    The contour argument was pioneered by Peierls in 1936 [8]
to demonstrate that the two dimensional Ising model does exhibit
phase coexistence at low temperature. The original argument
benefited from the particular symmetries of the Ising model.   The
adaptation of the method to the treatment of non-symmetric models
is not trivial, and was developed by Pirogov and Sinai [9], [13]
(see also [1]-[7],[15]). A particularly enlightening alternative
version of the argument was put forward by Zahradnik [14].

    In the Pirogov-Sinai (PS) theory  configurations can be
described by contours which satisfy Peierls condition. This theory
provides tools for a very detailed knowledge of the structure of
Gibbs measures in a region in the relevant parameters space (see
e.g. [13]). The PS theory is a low temperature expansion which
enables to control the entropic fluctuations from the ground
states, its natural setup being the lattice systems. But the
theory is not limited to such cases and it has been applied to a
great variety of situations, covering various types of phase
transitions. (see e.g. [3] for details).

    The main object of the theory is a family of contours defining a
configuration. In the original PS theory the ensemble of contours
has more complicated form. In particular, they do not have the
"contour-removal operation" (even for symmetric models) introduced
by Peierls.

 This paper presents a new definition of the contour on $Z^d$.
Contours defined here more convenient to prove the main theorem of
the PS theory for symmetric models. They allow as to use classical
Peierls argument (with the contour-removal operation). Such contours
for models on the Cayley tree were defined in [10]-[12].

The paper is organized as follows. In section 2 we give all
necessary definitions and check the Peierls condition. Section 3
devoted to definition and properties of new contours. In section 4
by the classical Peierls argument we show the existence of $s$
different (where $s$ is the number of ground states) Gibbs
measures.

\section{Definitions and Peierls condition}

{\it 2.1. Configuration space and the model.}  We consider the $d-$dimensional
($d\geq 2$) cubic lattice $Z^d$. The distance $d(x,y), \ \ x,y\in Z^d$ is defined by
$$ d(x,y)=\max_{1\leq i\leq d}|x_i-y_i|.$$

 For $A\subseteq Z^d$ a
spin {\it configuration} $\s_A$ on $A$ is  defined as a function
 $x\in A\to\s_A(x)\in\Phi=\{1,2,...,q\}$; the set of all configurations coincides with
$\Omega_A=\Phi^{A}$. We denote $\Om=\Om_{Z^d}$ and $\s=\s_{Z^d}.$
Also we define a {\it periodic configuration} as a configuration
$\s\in \Om$ which is invariant under a subgroup of shifts
$Z^d_*\subset Z^d$ of finite index. A configuration
 that is invariant with respect to all
shifts is called {\it translational-invariant}.

The energy of the configuration $\s\in \Om$ is given by the formal
Hamiltonian
$$
H(\s)=\sum\limits_{A\subset Z^d:\atop {\rm diam}(A)\leq r}I(\s_A)
\eqno (2.1)
$$
where $r\in N=\{1,2,...\},$ ${\rm diam}(A)=\max_{x,y\in A}d(x,y)$,
$I(\s_A): \Om_A\to R$ is a given translational-invariant potential.

 Denote by $M_r$ the set of all cubes of linear size $r$.

For $A\subset Z^d$ with ${\rm diam}(A)\leq r$ denote
$$n(A)=|\{b\in M_r: A\subset b\}|,$$
where $|B|$ stands for the number of elements of a set $B$.

The Hamiltonian (2.1) can be rewritten as
$$ H(\s)=\sum_{b\in M_r}U(\s_b), \eqno (2.2)$$
where $U(\s_b)=\sum_{A\subset b}{I(\s_A)\over n(A)}.$

For a finite domain $D\subset Z^d$ with the boundary condition
$\varphi_{D^c}$ given on its complement $D^c=Z^d\setminus D,$ the
conditional Hamiltonian is
$$H(\s_D\big
| \varphi_{D^c})=\sum_{b\in M_r:\atop b\cap D\ne \emptyset}U(\s_b),
\eqno(2.3)$$ where
$$ \s_b(x)=\left\{\begin{array}{ll}
\s(x) & \textrm{if \ \
$ x\in b\cap D$}\\
\varphi(x)& \textrm{if \ \ $x\in b\cap D^c.$}\\
\end{array}\right.
$$

{\it 2.2. The ground state.} A {\it ground state} of (2.2) is a
configuration $\varphi$ in $Z^d$ whose energy cannot be lowered by
changing $\varphi$ in some local region. We assume that (2.2) has
a finite number of translation-periodic (i.e. invariant under the
action of some subgroup of $Z^d$ of finite index) ground states.
By a standard trick of partitioning the lattice into disjoint
cubes $Q(x)$ centered at $x\in pZ^d$ with an appropriate $p$ and
enlarging the spin space from $\Phi$ to $\Phi^Q$ one can transform
the model above into a model on $pZ^d$ with only
translation-invariant or non periodic ground states. Such a
transformation was considered in [6]. Hence, without loss of
generality, we assume translation-invariance instead of
translational-periodic and we permute the spin so that the set of
ground states of the model be $GS=GS(H)=\{\s^{(i)}, i=1,2,..,s\},
1\leq s\leq q$ with $\s^{(i)}(x)=i$ for any $x\in Z^d.$

{\it 2.3. Gibbs measure.} We consider a standard sigma-algebra
${\cal B}$ of subsets of $\Om$ generated by cylinder subsets; all
probability measures are considered on $(\Om,{\cal B})$. A
probability measure $\mu$ is called a {\it Gibbs measure} (with
Hamiltonian $H$) if it satisfies the DLR equation: $\forall$
 finite $\Lambda\subset Z^d$ and $\sigma_\L\in\Om_{\L}$:
$$\mu\left(\left\{\sigma\in\Om:\;
\sigma\big|_{\L}=\sigma_\L\right\}\right)= \int_{\Om}\mu ({\rm
d}\omega)\nu^{\L}_\varphi (\sigma_\L),\eqno (2.4)$$ where
$\nu^{\L}_{\varphi}$ is the conditional probability:
$$ \nu^{\L}_{\varphi}(\sigma_\L)=\frac{1}{Z_{\L,\varphi}}\exp\;\left(-\beta H
\left(\sigma_\L\big |\,\varphi_{\L^c}\right)\right). \eqno (2.5)$$
Here $\beta={1\over T}, T>0-$ temperature and $Z_{\L, \varphi}$
stands for the partition function in $\L$, with the boundary
condition $\varphi$:
$$Z_{\L, \varphi}=
\sum_{{\widetilde\sigma}_\L\in\Om_{\L}} \exp\;\left(-\beta H
\left({\widetilde\sigma}_\L\,\big
|\,\varphi_{\L^c}\right)\right).\eqno (2.6)$$

{\it 2.4. The Peierls condition.}

Denote by ${\mathbf U}$ the collection of all possible values of
$U(\s_b)$ for any configuration $\s_b,$ $b\in M_r.$  Since
$r<+\infty$ we have $|{\mathbf U}|<+\infty.$ Put $U^{\min}=\min
\{U: U\in{\mathbf U}\}$ and
$$\l_0=\min\bigg\{{\mathbf U}\setminus \{U\in {\mathbf U}: U=U^{\min}\}\bigg\}-
U^{\min}.\eqno(2.7)$$

The important assumptions of this paper (see subsection 2.2) are
the following:

{\it Assumption A1.} The set of all ground states is $GS=\{\s^{(i)},
i=1,2,...,s\}, 1\leq s\leq q.$

{\it Assumption A2.} $\lambda_0>0$ i.e. ${\mathbf U}$ has at least
two distinct elements.

Let $P_s$ be the group of permutations on $\{1,...,s\}$. For $g\in
P_s, \ g=(g_1,...,g_s)$ and $\sigma\in \Omega$ define $g\sigma\in
\Omega$ by
$$g\sigma(x)=\left\{\begin{array}{ll}
g_{\sigma(x)}, \ \ {\rm if} \ \ \sigma(x)\in \{1,...,s\}\\
\sigma(x), \ \ {\rm if} \ \ \sigma(x)\in \{s+1,...,q\}.\\
\end{array}\right. $$

{\it Assumption A3.} Hamiltonian (2.1) is symmetric i.e.
$H(g\sigma)=H(\sigma)$ for any $g\in P_s$ and $\forall \sigma \in
\Omega$.

 {\it Remark.} If a configuration $\s$ satisfies
$$ U(\s_b)=U^{\min}\ \ \mbox{for}\ \ \forall b\in M_r \eqno(2.8)$$
then it is a ground state. Moreover for Hamiltonians on $Z^d$ it
is well known that a configuration is a ground state if and only
if the condition (2.8) is satisfied (see e.g. [13]).

The {\it relative Hamiltonian} is defined by
$$ H(\s,\varphi)=\sum_{b\in M_r}(U(\s_b)-U(\varphi_b)).$$
\vskip 0.4 truecm

{\bf Definition 2.1.} {\it Let $GS$ be the  set of all ground
states of the relative Hamiltonian $H$. A cube $b\in M_r$ is said
to be an {\sl improper} cube of the configuration $\s$ if $\s_b
\ne \f_b$ for any $\f\in GS.$ The union of the improper cubes of a
configuration $\s$ is called the {\sl boundary} of the
configuration and denoted by} $\d(\s).$

\vskip 0.4 truecm

{\bf Definition 2.2.} {\it The relative Hamiltonian $H$ with the
set of ground states $GS$  satisfies the Peierls condition if for
any $\f\in GS$ and any configuration $\s$ coinciding almost
everywhere with $\f$ (i.e. $|\{x\in Z^d: \s(x)\ne\f(x)\}|<\infty$)
$$H(\s,\f)\geq \lambda |\d(\s)|,$$
where $\l$ is a positive constant which does not depend on $\s$,
and $|\d(\s)|$ is the number of cubes in} $\d(\s).$

\vskip 0.4 truecm

{\bf Proposition 2.3.} {\it If assumptions A1-A2 are satisfied
 then the Peierls condition holds.}

\vskip 0.5 truecm

{\it Proof.} Suppose $\s$ coincides almost everywhere with  a ground
state $\f\in GS$ then we have $U(\s_b)- U^{\min}\geq \l_0$ for any
$b\in \d(\s)$ since $\f$ is a ground state. Thus
$$H(\s,\f)=\sum_{b\in M_r}(U(\s_b)-U(\f_b))=
\sum_{b\in \d(\s)}(U(\s_b)-U^{\min})\geq \l_0|\d(\s)|.$$
Therefore, the Peierls condition is satisfied for $\lambda=\l_0$.
The proposition is proved.

\vskip 0.4 truecm

\section{Contours}

Before giving our new contours let us recall the definition of the
contour defined in original Pirogov-Sinai theory (see [13]).

{\bf Definition 3.1.} Pair $\Gamma=(M, \s(M))$, (where $M={\rm
supp} \s(M)$ is a finite connected subset of $Z^d$), is called
contour of the configuration $\s$, if $M$ is a component (maximal
connected set) of the boundary $\d(\s)$.

Now we define our contours which are more convenient to use for the
symmetric models.

 Let $\L\subset Z^d$ be a finite set. Let $\s^{(i)}_{\L^c}\equiv i$, $i=1,...,s$ be
a constant configuration outside of $\L .$ For each $i$ we extend
the configuration $\s_\L$ inside $\L$ to the entire lattice  by the
$i$th constant configuration and denote it by $\s^{(i)}_\L$. The set
of such configurations we denote by $\Om^{(i)}_\L.$

For a given configuration $\s^{(i)}_\L\in \Om^{(i)}_\L$ denote
$V^{(j)}_\L\equiv V^{(j)}_\L(\s^{(i)}_\L)=\{t\in \L: \s^{(i)}_\L(t)=
j\}, j=1,...,q, j\ne i.$ Let $G_{\L,j}=(V^{(j)}_\L, L^{(j)}_\L) $ be
a graph such that
$$L^{(j)}_\L=\{(x,y)\in V^{(j)}_\L\times V^{(j)}_\L: d(x,y)=1\},\ \  j=1,...,q, j\ne i.$$
It is clear, that for a fixed $\L$ the graph $G_{\L,j}$ contains a
finite number $(=m)$ of maximal  connected subgraphs $G_{\L,j,p}$
i.e.
$$G_{\L,j}=\{G_{\L,j,1}, ..., G_{\L,j,m}\}, \ \ G_{\L,j,p}=(V^{(j)}_{\L,p},
L^{(j)}_{\L,p}), \ \ p=1,...,m; j\ne i.$$ Here $V^{(j)}_{\L,p}$
and $L^{(j)}_{\L,p}$ are the set of vertexes and  edges of
$G^{\L,j}_p$, respectively.

The (finite) graph $G_{\L,j,p}, j=1,...,q, j\ne i, p=1,...,m $ is
called {\it subcontour} of the configuration  $\s_\L^{(i)}.$

The set $V^{(j)}_{\L,p}, \ \ j=1,...,q, j\ne i, p=1,...,m$ is
called {\it interior}  of  $G_{\L,j,p}$, and is denoted by
Int$G_{\L,j,p}$. Note that the configuration $\s^{(i)}_\L$ takes
the same value $j$ at all points of the connected component
$G_{\L,j,p}$. This value is called {\it mark} of the subcontour.

For any two subcontours $T_1, T_2$ the distance ${\rm
dist}(T_1,T_2)$ is defined by
$${\rm dist}(T_1,T_2)=\min_{x\in {\rm Int}T_1\atop y\in {\rm Int}T_2}d(x,y),$$
where $d(x,y)$ is the distance between $x,y\in Z^d$ (see section
2.1).

\vskip 0.4 truecm

{\bf Definition 3.2.} {\it The subcontours $T_1,T_2$ are called
{\sl adjacent} if ${\rm dist}(T_1,T_2)\leq r.$ A set of
subcontours ${\cal A}$ is called {\sl connected} if for any two
subcontours $T_1,T_2\in {\cal A}$ there is a collection of
subcontours $T_1={\tilde T}_1, {\tilde T}_2,...,{\tilde T}_l=T_2$
in ${\cal A}$ such that for each $i=1,...,l-1$ the subcontours
${\tilde T}_i$ and ${\tilde T}_{i+1}$ are adjacent.}

\vskip 0.4 truecm

{\bf Definition 3.3.} {\it Any maximal connected set (component)
of subcontours (with given marks) is called {\sl contour} of the
configuration} $\s^{(i)}_\L.$

\vskip 0.4 truecm

For contour  $\g=\{T_p\}$ denote  ${\rm Int}\g=\cup_p{\rm Int}T_p.$

\vskip 0.4 truecm

{\it Remarks.} Our definition of a contour is different from the
Definition 3.1. Indeed: (i) our contour can be non connected
subgraph of $Z^d$, but the contours in original PS theory are
connected; (ii) By  our definition for any two contours $\g, \g'$ we
have ${\rm dist}(\g,\g')>r$. Thus our contours do not interact. This
means that for any $\s\in \Om$ there is no a cube $b\in \d(\s)$ with
$b\cap \g\ne\emptyset$ and $b\cap \g'\ne \emptyset.$ Such property
allows as to use a {\it contour-removal} operation. This operation
is similar to the one in ordinary Peierls argument [5]: Given a
family of contours defining a configuration $\s\in \Om^{(i)}_\L,$
the family obtained by omitting one of them is also the family of
contours of a (different) configuration in $\Om^{(i)}_\L.$ There is
an algorithm of the contour-removal operation to obtain a new
configuration as follows. Take the configuration $\s$ and change all
the spins in the interior of $\g$ (which must be removed) to value
$i.$ This makes $\g$ disappear, but leaves intact the other
contours. Contours defined in the Definition 3.1 may interact.
Therefore the Peierls argument is not directly applicable in that
approach.
 \vskip 0.4 truecm

In the sequel of the paper by contour we mean a contour defined by
Definition 3.3.  For a given (sub)contour $\g$ denote
$${\rm imp}\g=\{b\in \d: b\cap \g\ne
\emptyset\}, \ \ |\g|=|{\rm imp}\g|.$$

By the construction we have ${\rm imp}\g\cap {\rm imp}\g'=\emptyset$
for any contours $\g\ne\g'.$

For a given graph $G$ denote by $V(G)$ the set of its vertices.

Let us define a graph structure on $M_r$ as follows. Two cubes $b,
b'\in M_r$ are connected by an edge if $b\cap b'\ne\emptyset$.
Denote this graph by $G(M_r).$ Here the vertices of this graph are
elements (cubes) of $M_r.$ Note that the graph $G(M_r)$ is a
locally finite i.e. there is $k=k(d,r)<+\infty$  such that any
vertex of $G(M_r)$ has $k$ nearest neighbors. Thus Lemma 1.2 of
[2] can be reformulated as follows

\vskip 0.2 truecm

{\bf Lemma 3.4.} {\it Let ${\tilde N}_{n,G}(x)$ be the number of
connected subgraphs $G'\subset G(M_r)$ with $x\in V(G')$ and
$|V(G')|=n.$ Then}
$${\tilde N}_{n,G}(x)\leq (e k)^n.$$

 For
$x\in Z^d$ we will write $x\in \g$ if $x\in {\rm Int}\g.$

Denote $N_n(x)=|\{\g: x\in \g, |\g|=n\}|,$ where as before
$|\g|=|{\rm imp}\g|.$

\vskip 0.3 truecm

{\bf Lemma 3.5.} $N_n(x)\leq \frac{1}{2}(4e k)^n.$

\vskip 0.24 truecm

{\bf Proof.} Consider ${\rm imp}\g$ as a subgraph of $G(M_r)$. In
general ${\rm imp}\g$ may be non connected subgraph of the graph
$G(M_r)$.  Denote by $K_\g$ the minimal connected subgraph of
$G(M_r)$, which contains the contour $\g$. It is easy to see that
$$|V(K_\g)|\leq 2|{\rm imp}\g|=2|\g|. \eqno(3.1)$$
Using the estimation (3.1) and Lemma 3.4 we obtain

$$ N_n(x)\leq {2n \choose n} \tilde{N}_{2n, G}(x)\leq 2^{2n-1}(e k)^n
=\frac{1}{2}(4 e k)^n.$$
 The lemma is
proved.

\section{ Non-uniqueness of Gibbs measure}

For $A\subset Z^d$ denote
$$C(A)=\{b\in M_r: b\cap A\ne\emptyset\}.$$

For $\s_\L\in \Om^{(i)}_{\L}$ the conditional Hamiltonian (2.3) has
the form

$$H^{(i)}(\s_\L)\equiv
H(\s_\L\big | \s_{\L^c}=i)=\sum_{b\in M_r:\atop b\cap
\L\ne\emptyset}U(\s_{\L,b})=$$
$$\sum_{b\in \d(\s_\L)}(U(\s_{\L,b})-U^{\min})+|C(\L)|U^{\min},
\eqno(4.1)$$ where $\s_{\L,b}(x)=\s_\L(x)$ if $x\in \L\cap b$ and
$\s_{\L,b}(x)=i$ if $x\in \L^c\cap b$.

 The Gibbs measure on the
space $\Om^{(i)}_{\L}$ with boundary condition $\s^{(i)}$ is defined
as
$$\mu^{(i)}_{\L,\b}(\s_\L)={\bf Z}_{\L,i}^{-1}\exp(-\b H^{(i)}(\s_\L)), \eqno(4.2)$$
where ${\bf Z}_{\L,i}$ is the normalizing factor.

 Let us consider a sequence of sets  on $Z^d$
$$V_1\subset V_2\subset ... \subset V_n\subset ..., \ \ \cup V_n=Z^d,$$
and $s$ sequences of boundary conditions outside these sets:
$$\s^{(i)}_n\equiv i, n=1,2,..., i=1,...,s .$$
By very similar argument of proof of the lemma 9.2 in [7] one can
prove that each of $s$ sequences of measures $\{\m^{(i)}_{n,\b},
n=1,2,...\}, i=1,...,s$ contains a convergent subsequence.

We denote the corresponding limits by $\m^{(i)}_{\b}, i=1,...,s$.
 Our purpose is to show that for
 a sufficiently large $\b$ these measures are different.
\vskip 0.3 truecm

{\bf Lemma 4.1.} {\it Suppose assumptions A1, A2 are satisfied. Let
$\g$ be a fixed contour and $p_i(\g)=\mu^{(i)}_\b(\s_n\in \Om_{V_n}:
\g\in \d(\s_n)). $ Then
$$p_i(\g)\leq \exp\{-\b\l_0|\g|\}, \eqno(4.3)$$
where $\l_0$ is defined by formula (2.7).}
\vskip 0.2 truecm

{\it Proof.} Put $\Om_\g=\{\s_n\in \Om^{(i)}_{V_n}: \g\subset
\d(\s_n)\}$, $\Om_\g^0=\{\s_n: \g\cap \d=\emptyset\}$  and define a
(contour-removal)  map $\chi_\g:\Om_\g \to \Om_\g^0$ by

$$\chi_\g(\s_n)(x)=\left\{\begin{array}{ll}
i & \textrm{if \ \ $ x\in {\rm Int}\g$}\\
\s_n(x)& \textrm{if \ \ $x\notin {\rm Int}\g.$} \\
\end{array}\right.
$$
When $\g$ is fixed then the configuration on ${\rm Int}\g$ also
fixed. Therefore the map $\chi_\g$ is one-to-one map.
 For any $\s_n\in \Om^{(i)}_{V_n}$ we have
$$|\d(\s_n)|=|\d(\chi_\g(\s_n))|+|\g|.$$
Consequently, using (4.1) one finds
$$p_i(\g)={\sum_{\s_n\in \Om_\g}\exp\{-\b \sum_{b\in \d(\s_n)}(U(\s_{n,b})-U^{\min})\}\over
\sum_{{\tilde\s}_n}\exp\{-\b \sum_{b\in
\d({\tilde\s}_n)}(U({\tilde\s}_{n,b})-U^{\min})\}}\leq
$$
$${\sum_{\s_n\in \Om_\g}\exp\{-\b \sum_{b\in
\d(\s_n)}(U(\s_{n,b})-U^{\min})\}\over \sum_{{\tilde\s}_n\in
\Om^0_\g}\exp\{-\b \sum_{b\in
\d({\tilde\s}_n)}(U({\tilde\s}_{n,b})-U^{\min})\}}=$$
$${\sum_{\s_n\in \Om_\g}\exp\{-\b \sum_{b\in
\d(\s_n)}(U(\s_{n,b})-U^{\min})\}\over \sum_{{\tilde\s}_n\in
\Om_\g}\exp\{-\b \sum_{b\in
\d(\chi_\g({\tilde\s}_n))}(U(\chi_\g({\tilde\s}_{n,b}))-U^{\min})\}}.
\eqno(4.4)$$ Since $\s_{n,b}=\chi_\g(\s_{n,b}),$ for any $b\in
\d(\s_n)\setminus {\rm imp}\g$ we have
$$\sum_{b\in \d(\s_n)}(U(\s_{n,b})-U^{\min})=S_1+S_2, \eqno(4.5)$$ where
$S_1=\sum_{b\in \d(\chi_\g(\s_n))}(U(\s_{n,b})-U^{\min});$
$S_2=\sum_{b\in {\rm imp}\g}(U(\s_{n,b})-U^{\min}).$

By our construction $\g$ is a contour of $\d(\s_n)$ iff $\s_n(x)=i$
for any $x\in Z^d\setminus {\rm Int}\g$ with $d(x, {\rm Int}\g)<r$.
Consequently, ${\rm imp}\g$  does not depend on $\s_n\in\Om_\g.$ By
assumptions A1-A2 we have $U(\s_{n,b})-U^{\min}\geq \l_0>0,$ for any
$b\in {\rm imp}\g.$

Hence

$$S_2=\sum_{b\in {\rm imp}\g}(U(\s_{n,b})-U^{\min})\geq
\l_0|\g|, \ \ {\rm for \ any} \ \ \s_n\in \Om_\g. \eqno(4.6)$$

 Thus from (4.4)-(4.6) one gets (4.3). The lemma is
proved.

Now using Lemmas 3.5 and 4.1 by very similar argument of [11] one
can prove the following
 \vskip 0.3 truecm

{\bf Lemma 4.2.} {\it If assumptions A1-A3 are satisfied then for
fixed $x\in \L$ uniformly by $\L$ the following relation holds}
$$\mu^{(i)}_\b(\s_\L: \s_\L(x)=j)\to 0, j\ne i \ \ as \ \ \b\to \infty.$$

This lemma implies the main result, i.e. \vskip 0.2 truecm

{\bf Theorem 4.3.} {\it If A1-A3 are satisfied then for all
sufficiently large $\b$ there are at least $s$ (=number of ground
states) Gibbs measures for the Hamiltonian (2.2) on $Z^d$.}

 \vskip 0.2 truecm

 {\bf Acknowledgments.} The work supported by the SAGA Fund P77c
 of the Ministry of Science, Technology and Innovation (MOSTI) through the
 Academy of Sciences Malaysia. RUA thanks MOSTI and IIUM,
 for support and hospitality (in July-August 2007).

\vskip 0.3 truecm

{\bf References}

1. Biskup, M., Borgs, C., Chayes, J. T., Koteck\'y, R.: Partition
function zeros at first-order phase transitions: Pirogov-Sinai
theory. J. Stat. Phys. {\bf 116}, 97-155 (2004)

2. Borgs, C.: Statistical physics expansion methods in combinatorics
and computer science,  CBMS Lecture Series, Memphis 2003 (in
preparation).

3. Bovier, A., Merola, I., Presutti, E., Zahradnik, M.: On the Gibbs
phase rule in the Pirogov-Sinai regime. J. Stat. Phys. {\bf 114},
1235-1267 (2004)

4. Ganikhodjaev, N.,  Pah, C. H.: Phase diagrams of multicomponent
lattice models. Theor. Math. Phys. {\bf 149}, 244-251 (2006).

5. Fern$\acute{\textrm {a}}$ndez, R.: Contour ensembles and the
description of Gibbsian probability distributions at low
temperature. www.univ-rouen.fr/LMRS/persopage/Fernandez, 1998.

6. Lebowitz, J. L., Mazel, A. E.: On the uniqueness of Gibbs
states in the Pirogov-Sinai theory. Commun. Math. Phys. {\bf 189},
311-321 (1997)

7.  Minlos, R.A.:  Introduction to mathematical statistical
physics, University lecture series,  v.19, AMS,  2000.

8. Peierls, R.: On Ising model of ferro magnetism.  Proc.
Cambridge Phil. Soc. {\bf 32}, 477-481 (1936).

9. Pirogov, S.A., Sinai,Ya. G.: Phase diagrams of classical
lattice systems.I, II.  Theor. Math. Phys. {\bf 25}, 1185-1192
(1975); {\bf 26}, 39-49 (1976)

10. Rozikov, U.A.: An example of one-dimensional phase transition.
Siber. Adv. Math. {\bf 16}, 121-125 (2006)

11. Rozikov, U.A.: On $q-$ component models on Cayley tree:
contour method.  Lett. Math. Phys. {\bf 71}, 27-38 (2005)

12. Rozikov, U. A.: A constructive description of ground states
and Gibbs measures for Ising model with two-step interactions on
Cayley tree.  J. Stat. Phys. {\bf 122}, 217-235 (2006)

13. Sinai, Ya.G.:  Theory of phase transitions: Rigorous Results,
Oxford: Pergamon, 1982.

14. Zahradnik, M.: An alternate version of Pirogov-Sinai theory.
Commun. Math. Phys. {\bf 93}, 559-581 (1984)

15. Zahradnik, M.: A short course on the Pirogov-Sinai theory.
Rendiconti Math. Serie VII. {\bf 18}, 411-486 (1998)

\end{document}